\documentclass[12pt]{article} \textheight=24 true cm
\usepackage{graphicx}
\usepackage{amsmath}
\usepackage{amssymb}
\usepackage{color}
\usepackage[colorlinks]{hyperref}

\begin{document}
\begin{center}
{\Large\bf Spherically symmetric inhomogeneous model with
Chaplygin gas }\\[15 mm]
D. Panigrahi\footnote{ Sree Chaitanya College, Habra 743268, India
\emph{and also} Relativity and Cosmology Research Centre, Jadavpur
University, Kolkata - 700032, India , e-mail:
dibyendupanigrahi@yahoo.co.in }
  and S. Chatterjee\footnote{ IGNOU Convergence
Centre, New Alipore College, Kolkata - 700053, India \emph{and
also} Relativity and Cosmology Research Centre, Jadavpur
University,
Kolkata - 700032, India, e-mail : chat\_sujit1@yahoo.com\\Correspondence to : S. Chatterjee} \\[10mm]

\end{center}

\begin{abstract}

We investigate the late time acceleration with a Chaplygin type of
gas in spherically symmetric inhomogeneous model. At the early
phase we get  Einstien-deSitter type of solution generalised to
inhomogeneous spacetime. But at late stage of the evolution our
solutions admit the accelerating nature of the universe. For a
large scale factor our model behaves like a $\Lambda CDM$ model.
We calculate the deceleration parameter for this anisotropic
model, which, unlike its homogeneous counterpart, shows that the
\emph{flip} is not synchronous occurring early at the outer
shells. This is in line with other physical processes in any
inhomogeneous models. Depending upon initial conditions our
solution also gives bouncing universe. In the absence of
inhomogeneity our solution reduces to wellknown solutions in
homogeneous case. We have also calculated the effective
deceleration parameter in terms of Hubble parameter. The whole
situation is later discussed with the help of wellknown
Raychaudhury equation and the results are compared with the
previous case. This work is an extension of our recent
communication where an attempt was made to see if the presence of
extra dimensions and/or inhomogeneity can trigger an inflation in
a matter dominated Lemaitre Tolman Bondi model.
\end{abstract}

KEYWORDS : cosmology;  accelerating universe; inhomogeneity;\\
 PACS :   04.20,  04.50 +h
\bigskip
\section{ Introduction}
\vspace{0.1 cm}
 Following the high redshift supernovae data in the last
decade~\cite{riess} we know that when interpreted within the
framework of the standard FRW type of universe (homogeneous and
isotropic) we are left with the only alternative that the universe
is now going through an accelerated expansion with baryonic matter
contributing only five percent of the total budget. Later data
from CMBR studies~\cite{spergel} further corroborate this
conclusion which has led a vast chunk of cosmology community
(\cite{star} and references therein) to embark on a quest to
explain the cause of the acceleration. The teething problem now
confronting researchers is the identification of the mechanism
that triggered the late inflation. Workers in this field are
broadly divided into two groups - either modification of the
original general theory of relativity or introduction of any
mysterious fluid in the form of an evolving cosmological constant
or a quintessential type of scalar field. But as discussed
extensively in the literature (we are sparing the readers here to
repeat once again those arguments) both the alternatives face
serious theoretical problems. In this context one important thing
should not escape our attention. One intriguing fact in the
framework of the standard FRW model is that the accelerating phase
coincides with the period in which inhomogeneities in the matter
distribution at length scales $<¡« 10$ Mpc become significant so
that the Universe can no longer be approximated as homogeneous  at
these scales. One should also note that homogeneity and isotropy
of the geometry are not essential ingredients to establish a
number of relevant results in relativistic cosmology. One need not
be too sacrosanct about these concepts so as to sacrifice basic
physics (energy conditions, for example) in relativistic
cosmology. Conversely, if the universe is not \emph{apriori}
assumed to be homogeneous and isotropic, the observational data do
not necessarily imply an accelerating expansion of the universe,
or even if the cosmic expansion is accelerating it does not
necessarily point to an existence of a dark energy. Thus to
account for the observational data without introducing the concept
of dark energy, varied arguments regarding the effects of
\emph{inhomogeneities} have been made and naturally a vast
community of cosmologists have embarked upon a sort of  `mission'
to explain (sometimes with conflicting claims) the observational
findings within inhomogeneous models. The immediate generalisation
of FRW spacetime is the wellknown LTB model~\cite{ltb} which is
also spherically symmetric but the spacetime is inhomogeneous.
However, the assumption of spherical symmetry requires a centre of
the universe so that the observer be located not too far from the
centre to avoid undetected large anisotropy (a detailed study of
LTB and allied cosmologies and its relevance to current
astrophysical issues may be found in~\cite{kras}). The
\emph{sojourn} to the inhomogeneous path has a chequered history.
\emph{Naively }speaking there are two such arguments. One is that
the apparent acceleration of the cosmology can be regarded as a
result of an almost spherically symmetric but inhomogeneous
peculiar velocity field, assuming that we are located at the
vicinity of the symmetry centre~\cite{igu, van}. With this
argument the acceleration of the cosmic volume expansion is not
necessary. The other argument is that the  acceleration of the
universe is a physical reality and results from the backreaction
effects due to the inhomogeneities in the background FRW
universe~\cite{kolb,dp}. This idea is later supplemented by Carter
\emph{et al}~\cite{car} where the observed universe is assumed to
be  an underdense bubble in an Einstein-de Sitter universe and it
was shown that from observational point of view their results
become very similar to the predictions of $\Lambda CDM$ model.
However, in a recent communication Bolejko and
Andersson~\cite{bol} have calculated the volume deceleration based
on Buchert averaging scheme and back reaction in some LTB models
and have shown that for realistic cases the deceleration parameter
turns out to be positive. At this stage a very brief mention of
the formalism may not be out of place. The difference between the
evolution of homogeneous models and an inhomogeneous universe is
caused by backreaction effects, due to non linearity of Einstein
equations such that the solutions for a homogeneous matter
distribution leads in principle to a different description of the
universe than an average of an inhomogeneous solution to the exact
Einstein equations. So either we have to fall upon on exact
solutions or invoke averaging the backreaction terms. If simply a
volume averaging is considered then such an attempt leads to
Buchert equation. The Buchert equations are very similar to
Friedmann equations except for the backreaction term, which is, in
general nonvanishing if inhomogeneity is present(for a lucid
review of the averaging scheme the reader is referred to
~\cite{buch, rasa}). Moreover, the validity of the perturbative
ansatz is questionable in that the claimed acceleration is later
shown to be due to the result of extrapolation of a specific
solution to a regime where both the perturbative expansion breaks
down and  the constraints are violated \cite{giovannini}. On the
otherhand Kai \emph{et al} ~\cite{kai} showed that if one
hypothesizes the coexistence of expanding and contracting regions
in space, the speed of the cosmic volume expansion can be
accelerated. These models are constructed by replacing the
spherical domains from the Einstein-deSitter model with a LTB dust
sphere having the same gravitational mass. Another interesting
suggestion has recently come from the works of Wiltshire \emph{et
al}~\cite{wil} where a timescape cosmology has been proposed as a
viable alternative to homogeneous cosmologies with dark energy. It
realises cosmic acceleration as an apparent effect that arises in
calibrating average cosmological parameters in the presence of
spatial curvature and gravitational energy gradients that grow
large with the growth of inhomogeneities at late epochs. The model
is based on an exact solution to a Buchert average of the Einstein
equations with backreaticon. Some people attempted to look into
the problem from a purely geometric point of view - an approach
more in line with Einstein's spirits. For example, Panigrahi
\emph{et al}~\cite{sc} have recently toyed with the idea of
dimension driven acceleration in a number of publications, where
the extra terms coming from the higher dimensions create a sort of
back reaction to drive inflation. Good thing about it is that one
need not have to invent any exotic, unphysical matter field in
this case. In an interesting contribution Wanas~\cite{wanas}
introduced torsion to explain late acceleration. It is shown that
torsion generates a new type of energy to be called torsion energy
which is repulsive in nature, thus mimicing a quintessential type
of field. While torsion inspired inflation has several desirable
features(for example, geometrical origin) the problem with Wanas'
model is that the geometry no longer remains Riemannian.

 \vspace{0.1 cm} Again we know~\cite{book} that in a matter dominated nonrotating model where
particles interacting with one another move along geodesic lines
it is always possible to define a coordinate system which is at
once synchronous ($g_{00}=1$) and comoving. With this input Hirata
and Seljak ~\cite{hirata} claimed to have proved from Ray
Chaudhuri equation ~\cite{ray} that in a perfect fluid
cosmological model that is geodesic, rotation-free and obeys the
strong energy condition $(\rho + 3p)\geq 0$, a certain
generalisation of the deceleration parameter, $ q_{4}$ must be
always non-negative. But even with the perturbation considered by
Kolb \emph{et al}~\cite{kolb} the vorticity vanishes and
consequently Kolb's claim is flawed. On the other hand, Iguchi
\emph{et al} ~\cite{igu} did obtain simulated acceleration in
Lemaitre – Tolman (L–T) models with $\Lambda = 0$ that obey the
conditions set by Hirata \emph{et al}, which subsequently led
Vanderveld \emph{et al}~\cite{van} to draw attention to this
apparent contradiction between these two conclusions and to
suggest  that L–T models that simulate accelerated expansion also
contain a weak singularity, and in that case the derivation of HS
breaks down. In addition to this, there are other singularities
that tend to arise in L–T models, and Vanderveld \emph{et
al}~\cite{van} have failed to find any singularity-free models
that agree with observations. However in a pioneering contribution
Krasinski \emph{et al}~\cite{kra} neatly summed up the apparent
controversies as also the claims and counter claims of different
workers in this field and showed that the so called weak
singularity is not a singularity at all, while the other types of
singularities like shell crossing or shell focussing, generic to
all inhomogeneous collapse may be taken care of with suitably
chosen arbitrary functions appearing in the theory. Moreover, one
should point out at this stage that unlike the homogeneous case it
is always difficult, if not a little ambiguous to define uniquely
a deceleration parameter for inhomogeneous models. In this context
Krasinski \emph{et al} ~\cite{kra} also showed that
Hirata-Seljak's formulation of $q_{4}$ is wrong, based on
inadmissible averaging of the redshifts over directions and with
the averaging dropped one gets correct signature of the
deceleration parameter which may be both positive or negative. On
the other hand Hansson \emph{et al}~\cite{hansson} argued that
when taking the real, inhomogeneous and anisotropic matter
distribution in the semi-local universe into account, there may be
no need to postulate an accelerating expansion of the universe at
all despite recent type Ia supernova data. In fact inhomogeneous
structure formation may alleviate need for accelerating universe.

 \vspace{0.1 cm}

In the present work our goal is completely different. In a recent
communication one of us~\cite{skc} examined the possibility if the
presence of inhomogeneity or extra dimensions separately or
jointly in an LTB model can achieve late acceleration without the
aid of any extraneous scalar field. We found that while dimensions
have no perceptible effect on nature of evolution the radial or
angular acceleration is possible even in pure dust distribution if
any of them decelerates fast enough in LTB model. The present work
is an extension over that in the sense that we here introduce a
Chaplygin gas as input. The lack of information regarding the
provenance of dark matter and dark energy allows for speculation
with economical and aesthetic idea that a single component acted
in fact as both dark matter and dark energy. The unification of
these two components has risen considerable theoretical interest,
because on the one hand the model building has become considerably
simpler and on the other hand such unification implies the
existence of an era during which the energy densities of dark
matter and dark energy are remarkably similar. Moreover, at
present it is unclear whether the backreaction effects of
inhomogeneities can actually accelerate the cosmic volume
expansion.

 \vspace{0.1 cm}
One possible way to achieve that unification is through a
particular k-essence fluid, the Chaplygin gas with the exotic
equation of state. While literature abounds with work on Chaplygin
gas in FRW models (\cite{kam} and references therein), barring a
few~\cite{bil} we are not aware of works of similar kind directed
to inhomogeneous spacetime. However, relevant to mention that
Gorini \emph{et al}~\cite{gor} have recently discussed a
Tolman-Oppenheimer-Volkoff type of bounded distribution in
presence of a generalised Chaplygin gas. Here we discuss the
evolution of a spherically symmetric inhomogeneous model with a
Chaplygin type of matter field and get the interesting result that
an initially decelerating phase transits to  a late accelerating
one in line with the current observational results. We compared
our findings with those obtained via RayChaudhury equation also
and get identical results. Organisation of our paper is as
follows. In section 2 we  solve the field equations for our
inhomogeneous spacetime with Chaplygin gas as matter field and
have addressed the problems for both early and late time inflation
assuming a flat 3 space under different subtitles. While our
solutions are amenable to both early deceleration and late
acceleration we  get an interesting result that under suitable
initial conditions the model also admits a bouncing type of
universe avoiding the big crunch. In line with inhomogeneous
collapse the bounce occurs at different instants for different
shells unlike the synchronous homogeneous case. In section 3 we
compared our findings with the conclusions coming also from the
RayChaudhury equation. The paper ends with some concluding remarks
in section 4.

\section{ Field Equation}

\begin{equation}\label{a}
  ds^{2}= dt^{2}- X^{2}~dr^{2}-R^{2}(r,t)~dX^{2}
\end{equation}
where $dX^{2}$ represents a 2-sphere with

\begin{equation}
dX^{2}=d\theta^{2}+\sin^{2}\theta d\phi^{2}
\end{equation}

and the scale factor, $R(r,t)$  depends both on space and radial
coordinates $(r,t)$ respectively. A prime overhead denotes
${\partial}/{\partial}r$ and a  dot  denotes
${\partial}/{\partial}t$. As pointed out in the last section the
gauge, $g_{00}=1$ follows for a spherically symmetric,
irrotational inhomogeneous system only when, $p=0$. But in our
case, $p=0$ only under extremal condition and not all through its
evolution. So the \emph{ansatz}, $g_{00}=1$ should be treated as
an additional assumption to simplify field equations~\cite{sc1}.

 \vspace{0.1 cm}
 A comoving coordinate system is taken
such that $ u^{0}=1, u^{i}= 0 ~(i = 0,1, 2,3)$ and $g^{\mu
\nu}u_{\mu}u_{\nu}= 1$ where $u_{i}$ is the 4- velocity. The
energy momentum tensor for a dust distribution in the above
defined coordinates is given by

\begin{equation}
T^{\mu}_{\nu} = (\rho + p)\delta_{0}^{\mu}\delta_{\nu}^{0} -
p\delta_{\nu}^{\mu}
\end{equation}

where $\rho(r,t)$ is the matter density  and $p(r,t)$ is the
pressure. The fluid consists of successive shells marked by $r$,
whose local density ñ is time-dependent. The function $R(r, t)$
describes the location of the shell marked by $r$ at the time $t$.
Through an appropriate rescaling it can be chosen to satisfy the
gauge

\begin{equation}
 R(0, r) = r
 \end{equation}

The independent field equations for the metric (1) and the energy
momentum tensor (3) are given by

\begin{equation}
G^{0}_{0}= \frac{2\dot{X}\dot{R}}{XR}+ \frac{1 +
\dot{R}^{^{2}}}{R^{2}}-\frac{1}{X^{2}}\left(\frac{2R''}{R}+
\frac{R'^{2}}{R^{^{2}}}- 2\frac{X'R'}{XR}\right)= \rho
\end{equation}

\begin{equation}
G_{1}^{1} = 2\frac{\ddot{R}}{R}+  \frac{1 +
\dot{R}^{^{2}}}{R^{2}}- \frac{R'^{2}}{R^{^{2}}X^{2}}= -p
\end{equation}

\begin{equation}
G_{2}^{2}= G_{3}^{3}= \frac{\ddot{X}}{X}+\frac{\ddot{R}}{R}+
\frac{\dot{X}\dot{R}}{XR}- \frac{1}{X^{2}}\left(\frac{R''}{R}-
\frac{X'R'}{XR}\right)= -p
\end{equation}

\begin{equation}
G_{1}^{0}= -2\left( \frac{\dot{R'}}{R}-
\frac{R'}{R}\frac{\dot{X}}{X}\right)= 0
\end{equation}

Solving $G_{01}=0$ equation we get

\begin{equation} X(r,t) =
\frac{R'}{f(r)}
\end{equation}

where $f(r)$ is an arbitrary function of r.

 \vspace{0.2 cm}
Since the WMAP data~\cite{rasa}  shows that the universe is
spatially flat to within a few percent we can take $f=1$ such that
the field equations finally reduce to the following two
independent equations as

\begin{equation}
\frac{\dot{R}^{2}}{R^{2}}+ 2 \frac{\dot{R'}}{R'}\frac{\dot{R}}{R}=
\rho
\end{equation}

\begin{equation}
2\frac{\ddot{R}}{R}+ \frac{\dot{R}^{2}}{R^{2}}= -p
\end{equation}

From the the Bianchi identity we get for the inhomogeneous model
the conservation law

\begin{equation}
\nabla_{\nu}T^{\mu \nu}= 0
\end{equation}

which, in turn, yields

\begin{equation}
\delta_{\mu}p + \frac{1}{\sqrt-g}\delta_{\nu}\left[\sqrt-g(\rho +
p)u^{\mu}u^{\nu}\right] + \Gamma^{\mu}_{\nu
\lambda}u^{\nu}u^{\lambda}=0
\end{equation}

For $\Gamma^{\mu}_{00}=0$ and $\sqrt-g = X(r,t) R^{2}(r,t)
sin\theta$, we obtain

\begin{equation}
\frac{d\rho}{dt} +
\frac{1}{XR^{2}}\frac{d}{dt}\left(XR^{2}\right)(\rho +p)=0
\end{equation}

At this stage we assume that we are dealing with a Chaplygin type
of gas obeying an equation of state

\begin{equation}
p = -\frac{A}{\rho}
\end{equation}

where $A$ is a positive constant. It was first introduced as a
cosmological fluid unifying dark matter and dark energy by
Kamenshchik \emph{et al}~\cite {kam} and since has been widely
studied in this context. Moreover it has found applications in
particle physics via string theory~\cite{string} and its
supersymmetric extension~\cite{symm}. With this input we finally
get

\begin{equation}
\dot{\rho} +
\frac{1}{XR^{2}}\frac{d}{dt}\left(XR^{2}\right)\left(\rho -
\frac{A}{\rho}\right) = 0
\end{equation}

which integrates to

\begin{equation}
\rho = \left[ A + \frac{C(r)}{X^{2}R^{4}}\right]^{\frac{1}{2}}
\end{equation}

This becomes via equation(9)

\begin{equation}
\rho = \left[ A + \frac{C(r)}{R'^{2}R^{4}}\right]^{\frac{1}{2}}
\end{equation}

Plugging in the expression of $\rho$ from equation (10) we finally
get

\begin{equation}
\frac{\dot{R}^{2}}{R^{2}}+ 2 \frac{\dot{R'}}{R'}\frac{\dot{R}}{R}
= \left[ A + \frac{C(r)}{R'^{2}R^{4}}\right]^{\frac{1}{2}}
\end{equation}

\begin{figure}[ht]

\begin{center}
    \includegraphics[width=14 cm]{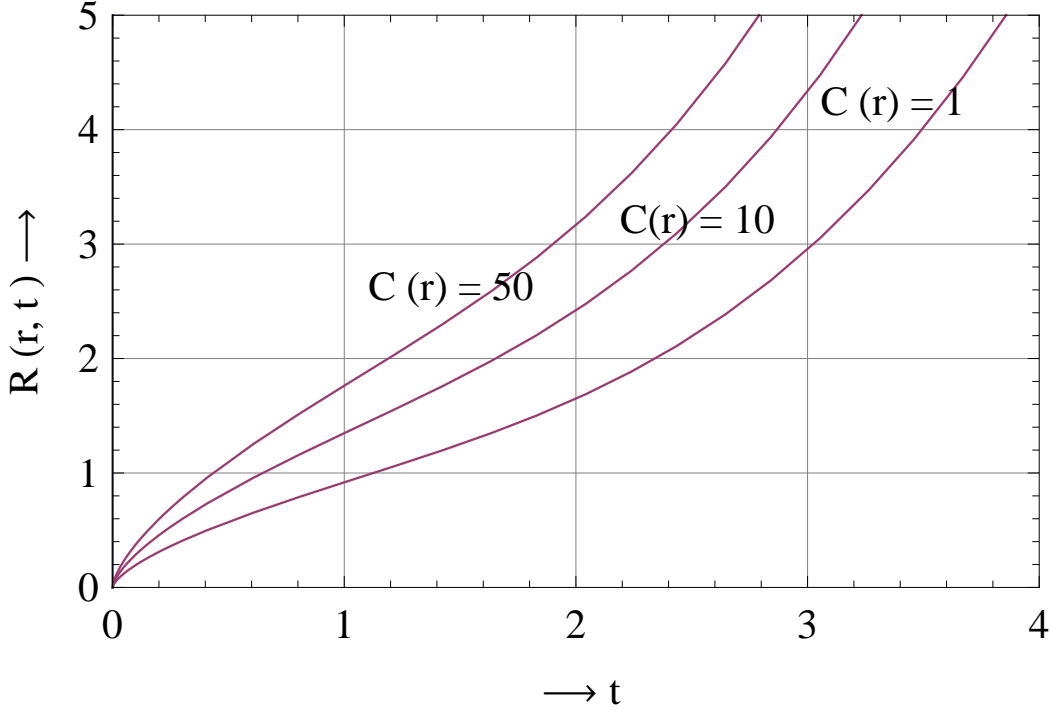}
  \caption{
  \small\emph{The variation of $R(r,t)$ and $t$  for different
  values of $C(r)$  is shown. The graphs clearly show that acceleration increases for greater C(r) i.e., greater inhomogeneity.
   }\label{1}
    }
\end{center}
\end{figure}

This is the key equation in most accelerating models dealing with
a Chaplygin gas in homogeneous models except that $C(r)$ is now
not a true constant but depends on space for inhomogeneous
expansion. Moreover the above expression is not amenable to any
closed form analytic solution but integration results in a
hypergeometric series. Figure-1 shows that acceleration depends on
$C(r)$, which represents the inhomogeneity. So $C(r)$ may be the
measure of inhomogeneity in our case.
 \vspace{0.1 cm}

\textbf{CASE~A:} At the early stage of the cosmological evolution
when the scale factor $R (r,t)$ is relatively small the second
term of the last equation (19) dominates and we get a sort of dust
dominated universe for a particular value of $C(r) =
\frac{16}{9}r^{4}$ yielding

\begin{equation}
R(r,t)= r\left[ t+ t_{0}(r)\right]^{\frac{2}{3}}
\end{equation}

It has not also escaped our notice that with this scale factor we
get a vanishing pressure when used in equation (11). Moreover for
isotropic expansion ($X'=R$) we get $\rho\sim\frac{1}{R^{3}}$ as
in FRW universe. Relevant to point out that the expression (20) is
not exactly Tolman-Bondi like and our line element reduces to

\begin{equation}
ds^{2}= dt^{2} - r^{2}\left[ t +
t_{0}(r)\right]^{\frac{4}{3}}(dr^{2} + r^{2}d\Omega^{2})
\end{equation}
If, at this stage, we assume that $t_{0}(r)$ vanishes or losing
its space dependence becomes a \emph{true} constant (in that case
a time translation is required) then we get

\begin{equation}
ds^{2}= dt^{2} - r^{2} t^{\frac{4}{3}}(dr^{2} + r^{2}d\Omega^{2})
\end{equation}

This is a new solution and may be termed as the generalised
Einstein- deSitter metric for the inhomogeneous spacetime. In the
analogous homogeneous case for zero pressure dust with vanishing
spatial curvature($k=0$)  we get for FRW metric the wellknown
Einstein-deSitter metric as($R\sim t^{\frac{2}{3}}$). From
equation (18) we get the expression of density as
\begin{equation}
\rho(r,t)\approx \frac{\sqrt C(r)}{R'R^{2}}= \frac{ 4}{3r \left[t+
t_{0}(r)\right]\left[\frac{t + t_{0}(r)}{r}+
\frac{2}{3}t'_{0}\right]}
\end{equation}

 \vspace{0.1 cm}
 \textbf{CASE~B :} ($R(r,t)$ is very large)

\textbf{Type - 1:}

 In the late stage of evolution the second term of the RHS of the
 equation (19) vanishes and we get

\begin{equation}
\frac{\dot{R}^{2}}{R^{2}}+ 2 \frac{\dot{R'}}{R'}\frac{\dot{R}}{R}
= \sqrt A
\end{equation}

This yields a solution

\begin{equation}
R (r,t) =  R_{0} \exp\left[{\frac{A^{\frac{1}{4}}}{\sqrt
3}(r+t)}\right]
\end{equation}

This is the wellknown de Sitter type of solution generalised to
inhomogeneous space. $A^{\frac{1}{2}}$ behaves as $\Lambda$, the
cosmological constant. However, if we make a radial coordinate
transformation

\begin{equation}
\bar{r}= R_{0}\exp \left[{\frac{A^{\frac{1}{4}}}{\sqrt
3}(r+t)}\right]
\end{equation}

 the metric reduces to

\begin{equation}
ds^{2}= dt^{2} - \exp
\left({2~\sqrt\frac{A^{\frac{1}{2}}}{3}~t}\right) d\bar{r}^{2} -
\bar{r}^{2}\exp \left({2~\sqrt\frac{A^{\frac{1}{2}}}{3}~t}\right)
~d\Omega^{2}
\end{equation}

\textbf{Type - 2:}
 \vspace{0.1 cm}
The volume expansion rate for our  metric is  defined through the
4-velocity of the fluid, $u^{a}$ as

\begin{equation}
3H = u^{a}_{; a}=u_{a ; b}~ g^{ab}=u_{a ; b}~ h^{ab}
\end{equation}

where

\begin{equation}
h^{ab}= g^{ab}+ u^{a}u^{b}
\end{equation}

As commented earlier in the discussion while the definition works
perfectly well for a FRW like homogeneous distribution of matter
it is always a bit ambiguous to define the deceleration parameter
of an inhomogeneous anisotropic model because the relation (28)
does not take into account the directional preference of the
matric. For example, Tolman-Bondi has a preferred direction, being
the radial one. We can still give an operational definition to the
average \emph{volume acceleration} of our model. For inhomogeneous
model the directional preference need to be emphasized in the
expression for expansion.  We define a projection tensor $t^{ab}$
that projects every quantity perpendicularly to the preferred
spacelike direction $s^{a}$ (and of course the timelike vector
field, $u^{a}$) such that

\begin{equation}
t^{ab} = g^{ab}+ u^{a}u^{b} - s^{a}s^{b}= h^{ab}- s^{a} s^{b}
\end{equation}

For our metric (1),

\begin{equation} p^{a}= \frac{\sqrt{1+
f(r)}}{R'}\bigtriangledown\end{equation}

and the tensor projects every physical quantity in a direction
$\bot$ to $s^{a}$. One can now define the invariant expansion
rates as

\begin{equation}
H_{r}= u_{a ; b} s^{a}s^{b} = \frac{\dot{R'}}{R'}
\end{equation}

\begin{equation}
H_{\bot}= \frac{1}{n}u_{a ; b}t^{ab}= \frac{\dot{R}}{R}
\end{equation}

so that

\begin{equation}
H = \frac{2}{3}H_{\bot} + \frac{1}{3}H_{r}
\end{equation}

Evidently the above definition gives a sort of averaging over the
various directions for our anisotropic model.
\\
If one relaxes the condition of any particular preferred direction
(like the radial one as in LTB model) one can explore the
definition of the Hubble parameter in a more transparent way
considering its directional dependence \cite{partovi} as follows:

 \vspace{0.1 cm}

\begin{equation}
H = \frac{1}{3}u^{a}_{_{a}} + \sigma_{a b}J^{a}J^{^{b}}
\end{equation}

where $\sigma_{a b}$ is the shear tensor and $J^{^{a}}$ a unit
vector pointing in the direction of observation. For an observer
located  away from the centre of the configuration it gives for
our LTB case

\begin{equation}
H = \frac{\dot{R}}{R} + \left(\frac{\dot{R'}}{R} -
\frac{\dot{R}}{R}\right) cos^{2}\theta
\end{equation}

 where $\theta$ is the angle between the radial
direction through the observer and the direction of observation.
Naturally when the two directions coincide, $\theta=0$ we get
$H=H_{_{r}}$ and for $\theta=\frac{\pi}{2}$ it is $H =
H_{\theta}$. A definition of deceleration parameter in a preferred
direction can also be given in terms of the expansion of the
Luminosity distance $D_{L}$ in powers of redshift of the incoming
photons. For small $z$ one gets

\begin{equation}
q = - \dot{H}\frac{d^{2}D_{L}}{dz^{2}}+ 1
\end{equation}

For $\theta =0$ and $\theta=\pi/2$ the acceleration is
respectively

\begin{equation}
q_{r} = -
\left(\frac{R}{\dot{R'}}\right)^{2}\left[\frac{\ddot{R'}}{R}-\frac{\sqrt{1
+ f}} {\dot{R'}}\left(\frac{\dot{R'}}{R}\right)'\right]
\end{equation}

\begin{equation}
q_{\bot}= - \left(\frac{R}{\dot{R}}\right)^{2}\frac{\ddot{R}}{R}
\end{equation}

We shall subsequently see in section 3 that deceleration parameter
defined this way has an important difference from what we later
get in equation (56). Here the parameters do not depend solely on
local quantities as opposed to the acceleration parameter of (56).
For example we get via  field equation(10-11)

\begin{equation}
q_{\bot}= \frac{M(r)}{R^{3}}\frac{1}{H_{\bot}^{2}}
\end{equation}
where $M(r)$ is the mass of the fluid distribution upto the
comoving radial coordinate$r$. Thus the equation (40) tells us
that here the deceleration parameter $q_{\bot}$ depends on the
total mass function and not on the local energy density of (48).

 \vspace{0.2 cm}
One can look into the above expression of $q_{\perp}$ from a
different standpoint also to assume a particular form of
deceleration
  parameter as
\begin{equation}
q_{\perp} = - \frac{1}{H_{\perp}^{2}}\frac{\ddot{R}}{\dot{R}} =
\frac{a - R^{m}}{b + R^{m}}
\end{equation}
where $a, b$ and $m$ are constants. Straight forward integration
of equation (41) yields
 \begin{eqnarray}
 R(r,t) = R_{0} \mathrm{sinh}^{n}\omega (r + t)
 \end{eqnarray}
 where, $n= \frac{2}{m}$, $a = \left(R_{0}\right)^\frac{2}{n}\left(\frac{1}{n} -
 1\right)$ and $b = \left(R_{0}\right)^{\frac{2}{n}}$
such that we get from equation (41)
\begin{equation}
q_{\perp} = \frac{1-n~\mathrm{cosh}^{2}\omega (r +
t)}{n~\mathrm{cosh}^{2}\omega (r + t)}
\end{equation}
showing that the exponent $n $ determines the evolution of
$q_{\perp}$. A little inspection shows that \emph{(i)} $a < 0$,
\emph{i.e.}, $n
> 1$ gives acceleration, \emph{(ii)}  $a > 0$, \emph{i.e.}, $0< n
<1$ gives the desirable feature of \emph{flip}, although it is not
obvious from our analysis at what value of redshift this
\emph{flip} occurs.

 \vspace{0.2 cm}
For $n = \frac{2}{3}$, equation (24) is satisfied for $A =
\frac{16}{\sqrt{9}}\omega^{4}$  ~and in this case

\begin{equation}
q_{\perp} = \frac{3}{2}\mathrm{sech}^{2}\omega(r+t)-1
\end{equation}
Figure -2 shows  the variation of $q_{\perp}$ and $t$ for
different values of $r$. We have seen from the graph that
\emph{flip} ($t_{c}$) occurs early at greater value of $r$,
\emph{i.e.}, acceleration depends on inhomogeneity. The
\emph{flip} time $(t_{c})$ will be in this case
\begin{equation}
t_{c} = \frac{1}{\omega}\left[-r +
\mathrm{sech}^{-1}\left(\sqrt{\frac{2}{3}} \right)\right]
 \end{equation}

\begin{figure}[ht]
\begin{center}
  \includegraphics[width=10cm]{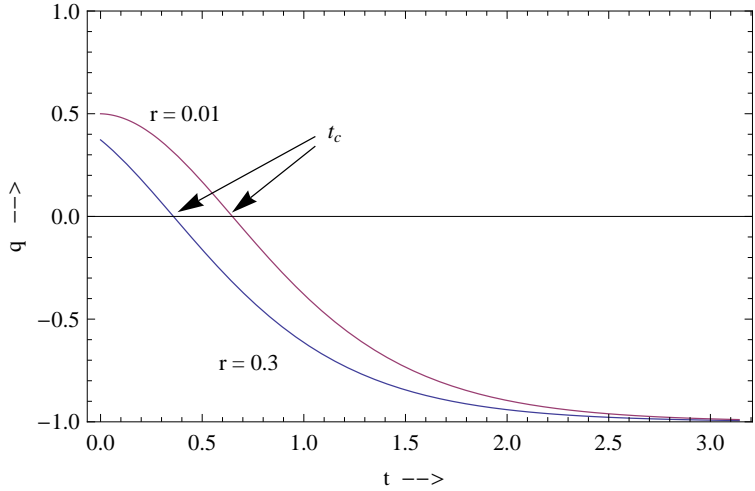}
  \caption{
  \small\emph{The variation of $q_{\perp}$ and $t$  for different
  values of $r$  is shown. The graphs clearly show that flip $(t_{c})$
  occurs early at greater value of r, i.e., acceleration increases
  for greater r i.e., greater inhomogeneity.
   }\label{1}  }
\end{center}
\end{figure}

\textbf{Type-3:}

 \vspace{0.2 cm}
 It also follows from equation (18) that for
the late universe ($R\sim \infty$)
\begin{equation}
\rho \simeq \sqrt A  + \frac{C(r)}{\sqrt{4A}}\frac{1}{R'^{2}R^{4}}
\end{equation}
\begin{equation}
p \simeq  -\sqrt A  + \frac{C(r)}{\sqrt{4A}}\frac{1}{R'^{2}R^{4}}
\end{equation}
This is a mixture of a Cosmological Constant $\sqrt{A}$ with a
type of matter obeying a `stiff fluid' equation of state. However
it should be pointed out that it is an inhomogeneous and
anisotropic generalization of the well known FLRW situation
characterized by $X (r,t) = R (r, t)$ where the quantities depend
on time only.

 Again as $R(r,t) \rightarrow \infty$, we asymptotically get $ p = - \rho$ from this Chaplygin
type of gas, which corresponds to an empty universe with
Cosmological constant $\sqrt{\frac{A}{3}}$.

\textbf{CASE~C :}

Now we are trying to solve the equation (19) using the method of
separation of variables. Let $R(r,t) = g(r) a(t)$. From equation
(19) we get
\begin{equation}
3 \frac{\dot{a}^2}{a^2} = \left( A +
\frac{B}{a^6}\right)^{\frac{1}{2}}
\end{equation}
where $B = \frac{C(r)}{g'^{2}g^4}$= Constant (say).

The equation (48) gives the hypergeometric solutions of $a(t)$
with $t$. The solution and other features are same like homogenous
case~\cite{kam,ujj} such as

 \vspace{0.1 cm}
\textbf{i) } When $A = 0$, we get pressureless equation of state.
Our solution reduces to FRW type and in this case $a(t) \sim
t^{\frac{2}{3}}$.

 \vspace{0.1 cm}
\textbf{ii) } At early stage of evolution, \emph{i.e.}, for small
value of $a(t)$, the equation (48) reduces to $3
\frac{\dot{a}^2}{a^2} =
 \frac{\sqrt{B}}{a^3} $ and we get $a(t) \sim
t^{\frac{2}{3}}$.

 \vspace{0.1 cm}
\textbf{iii) } At the late stage of evolution, i.e., $a(t)$ is
large in this case, the equation (48) becomes (neglecting higher
order terms )

\begin{equation}
3 \frac{\dot{a}^2}{a^2} =   \sqrt{A} + \frac{B}{\sqrt{4A}}a^{-6}
 \end{equation}
\begin{figure}[ht]

\begin{center}
  \includegraphics[width=10cm]{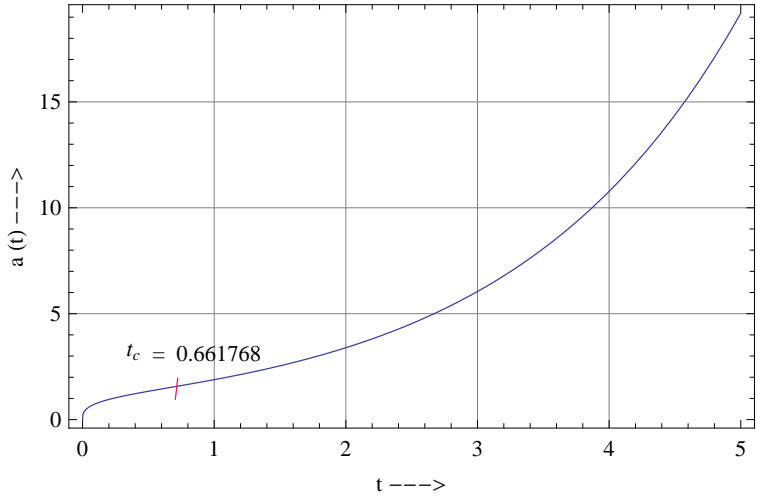}
  \caption{
  \small\emph{The variation of $a(t)$ vs $t$  is shown in this
  figure. Taking A=1 \& B=12. Flip occurs at $t = t_{c} = 0.661768$ }\label{1}
    }
\end{center}
\end{figure}
 Solving the equation (49) we get the solution,

\begin{equation}
a(t) =\sqrt{2} \frac{3^{\frac{1}{6}}}{\sqrt{A^{\frac{1}{6}}}}~e^{-
\frac{A^{\frac{1}{4}}}{\sqrt{3}}t}
  \left[ e^{2\sqrt{3}A^{\frac{1}{4}}t }  - \frac{B}{12
\sqrt{A}} \right]^{\frac{1}{3}}
 \end{equation}
 and also
\begin{equation}
g(r)^3 = \pm \frac{3}{\sqrt{B}} \int C(r) dr
 \end{equation}
 So
\begin{equation}
R(r,t) = \sqrt{2}
\frac{3^{\frac{1}{6}}}{\sqrt{A^{\frac{1}{6}}}}~e^{-\frac{A^{\frac{1}{4}}t}{\sqrt{3}}
} \left[\left( e^{2\sqrt{3}A^{\frac{1}{4}}t }  - \frac{B}{12
\sqrt{A}}\right)  \frac{3}{\sqrt{B}} \int C(r) dr
\right]^{\frac{1}{3}}
 \end{equation}
The nature of $R(r,t)$ with $t$ for a typical $r$ is shown in the
fig. - 3.

 \vspace{0.1 cm}
 In this
case, $R(r,t) = 0$ at  $t_{0} = \frac{1}{2
\sqrt{3}}A^{-\frac{1}{4}}\ln \left(\frac{B}{12 \sqrt{A}}\right)$.
 Here $\ddot{R} = 0$ at $t=\frac{\text{ln}\left[\sqrt{\frac{5
B}{12 \sqrt{A}}+\frac{B}{\sqrt{6} \sqrt{A}}}\right]}{\sqrt{3}
A^{1/4}}=t_{c}$, which is the \emph{filp} time.

\textbf{CASE~D :} (For Negative Integration Constant) \vspace{0.1
cm}

 Considering negative value of $C(r)$ presents interesting
possibilities~\cite{42} since
 in that case the energy density increases with scale factor mimicing the phantom
 dark energy model and finally ending up as a cosmological constant. We
get from equation(18)that for well behaved matter field the
condition $ R'^{2} R^{4}
> \frac{C(r)}{A}$, \emph{i.e.} need to be satisfied.  So a  minimal value of scale factor
exists for a typical value $r$, which is $R(r,t)_{min} =
\left[\frac{C(r)}{A}\right]^{\frac{1}{6}}$, pointing to a bouncing
universe at early times. We thus see that the Chaplygin gas model
interpolates between dust at small R and a cosmological constant
at large R, but choosing a negative value for C(r), this
quartessence idea lose. Following Barrow~\cite{bar} if we
reformulate the dynamics with a  scalar filed $\zeta$ and a
potential V  to mimic the Chaplygin cosmology, we see that a
negative value for B  dictates that we transform $\zeta = i\Psi$.
 In this case the expressions for the energy density and the pressure
 corresponding to the scalar field show that it represents a
 a phantom field. This implies that one can generate phantom-like
equation of state from an interacting generalized Chaplygin gas
dark energy model in LTB universe. This feature has been discussed
in the past~\cite{buch}in the context of an effective description
of inhomogeneous model evolving like a homogeneous solution
following an averaging technique and also in details
in~\cite{go,gor} . In our case the bounce is inhomogeneous in the
sense that each shell characterised by a constant radial
coordinate$r$ bounces at its own time. So the bounce is not
synchronous each shell sharing a local dynamics.

 \vspace{0.1 cm}
\section{ Raychaudhuri Equation :}

It may not be out of place to address the situation discussed in
the last section with the help of the well known Ray Chaudhuri
equation \cite{ray},  which in general holds for any cosmological
solution based on Einstein's gravitational field equations. The
Ray Chaudhuri equation is
\begin{equation}
\theta_{,\mu}v^{\mu} = \dot{v}^{\mu}_{; \mu} - 2(\sigma^{2}-
\omega^{2})-\frac{1}{3}~~ \theta^{2} + R_{\nu
\alpha}v^{\nu}v^{\alpha}
\end{equation}
where the terms have their usual significance. With matter field
expressed in terms of mass density and pressure Ray Chaudhury
equation is finally given by,
\begin{equation}\label{a}
  \dot{\theta}=-2(\sigma^{2}-\omega^{2})-\frac{1}{3}\theta^{2}-\frac{8\pi G}
  {2}\left(\rho+3p \right)
\end{equation}
in a  co moving reference frame. Here $p$ is the
 isotropic pressure.

  \vspace{0.1 cm}
 With the help of equation(34) we get an expression for effective
 deceleration parameter as
\begin{equation}\label{b}
 q= -\frac{\dot{H} + H^{2}}{H^{2}} = -1-3~\frac{\dot{\theta}}{\theta^{2}}
\end{equation}
 which allows us to write,
\begin{equation}\label{c}
  \theta^{2}q = 6 \sigma^{2}+ 12 \pi G \left( \rho
  + 3p \right)
\end{equation}

With the help of the equations (15), (18) \& (56) we finally get,

\begin{equation}\label{c}
  \theta^{2}q = 6 \sigma^{2}+ 12 \pi G \left[ -2A + \frac{C(r)}{R'^{2}
  R^{4}} \right]\left[ A + \frac{C(r)}{R'^{2} R^{4}} \right]^{-\frac{1}{2}}
\end{equation}
In our case the shear scalar evolves as
\begin{equation}
\sigma^{2} = \sigma_{\mu\nu}\sigma^{\mu \nu} = \frac{2}
{3}\left(\frac{\dot{R'}}{R'}- \frac{\dot{R}}{R}\right)^{2}
\end{equation}

\textbf{CASE~A : Early Stage:} At the early phase of this
evolution when the scale factor $R(r,t)$ is small enough the above
equation reduces to

\begin{equation}\label{c}
  \theta^{2}q = 6 \sigma^{2}+ 12 \pi G~ \frac{~~\left[C(r) \right]^{\frac{1}{2}}}{R'R^{2}}
\end{equation}

It follows from the equation (59) that $q$, the deceleration
factor is always positive. So  accelerated expansion is absent  in
this dust dominated phase though inhomogeneity is present here.
Interestingly this result is very similar to  the work of Alnes
\emph{et al }~\cite{alnes}.

 \vspace{0.1 cm}
 \textbf{CASE~B : Late Stage :}

 \vspace{0.1 cm}
 \textbf{Type - I:} If we consider
the late stage of evolution i.e., $R(r,t)$ is large enough in this
phase, the second term of the RHS of the equation (19) vanishes
and we get from equation (57),

\begin{equation}\label{c}
  \theta^{2}q = 6 \sigma^{2}- 24 \pi G \sqrt{A}
\end{equation}

At this stage if we consider the scale factor given by equation
(42) ($n = \frac{2}{3}$) the shear scalar becomes $\sigma^{2} =
\frac{8}{3} \omega^{2} \mathrm{cosech} ^{2} \left[2 \omega \left(
r+t \right) \right]$ \& $A = \frac{16}{9}\omega^{4}$. The equation
(57) reduces to

\begin{equation}\label{c}
  \theta^{2}q = 16 \omega^{2} \mathrm{cosech} ^{2} \left[2 \omega \left(
r+t \right) \right] - 32 \pi G \omega^{2}
\end{equation}
\begin{figure}[h]
\begin{center}
  \includegraphics[width=10cm]{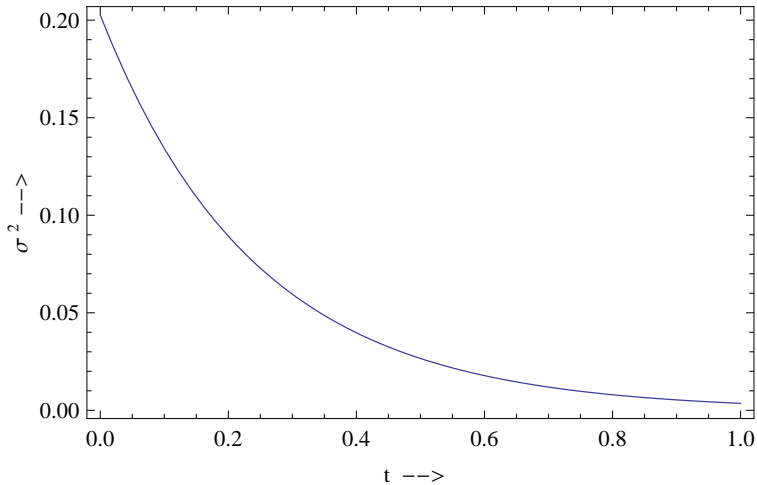}
  \caption{
  \small\emph{The variation of $\sigma^{2}$ vs $t$  is shown in this
  figure. Taking $\omega=1$ \& $r=1$.  }\label{1}
    }
\end{center}
\end{figure}

In figure - 4  shows $\sigma^{2}$ vs $t$ for a particular $r$. In
this graph we have seen that as $t$ increases $\sigma^{2}$
decreases, \emph{i.e.}, when $t \rightarrow \infty $, $\sigma^{2}
\rightarrow 0$. So  initially we get the decelerating universe and
after \emph{flip} it becomes accelerating in line with current
observational result (see equation (60)).

\textbf{Type - II:} Again if we consider first order approximation
of equation (57), neglecting higher order terms, we get
\begin{equation}\label{c}
  \theta^{2}q = 6 \sigma^{2}+ 24 \pi G A \left[ -A + \frac{C(r)}{R'^{2}
  R^{4}} \right]
\end{equation}
Let $R(r,t) = g(r)a(t)$, so in this case $\sigma = 0$ which
follows from the equation (58). Now the equation (62) reduces to
\begin{equation}\label{c}
  \theta^{2}q =  24 \pi G A \left[ -A + \frac{B}{a^{6}
  } \right]
\end{equation}

It follows from the equation (63) that flip occurs when $a(t) =
\left(\frac{B}{A} \right)^{\frac{1}{6}}$. Now $q < 0$, at $a(t)
> \left(\frac{B}{A}\right)^{\frac{1}{6}}$ i.e., acceleration takes place in this case.

Negative Constant : If $C(r) < 0$, then $B < 0$, the equation (63)
then becomes
\begin{equation}\label{c}
  \theta^{2}q = - 24 \pi G A \left[ A + \frac{B}{a^{6}
  } \right]
\end{equation}
From the above equation we have seen that always $q < 0$, which
means we get always accelerating universe.\\

\section{Concluding Remarks}
The present work may be looked upon as an extension of one of our
recent publications where we  examined the possibility in a higher
dimensional LTB model if the inclusion of extra space jointly with
inhomogeneity can induce late inflation in a dust model. While
total volume acceleration is ruled out we found that preferential
acceleration in radial direction  is possible if the angular
direction decelerates fast enough or \emph{vice versa}. Here we
have taken a Chaplygin type of gas as matter field to work out the
same problem in a 4D spacetime. While there is a proliferation of
work in the literature on homogeneous FRW model with Chaplygin gas
we have not much come across work of similar type in inhomogeneous
spacetime. Given the fact that it is difficult, if not a little
confusing to define uniquely a deceleration parameter in
inhomogeneous, anisotropic model we have nevertheless got the
following definitive results.\\
1. Aside from  space dependence the mathematical structure and its
followup in the section 2 is essentially similar to the works of
homogeneous spacetime except the appearance of the term, $C(r)$ in
equation(17), which unlike its homogeneous counterpart is not a
true constant but depends on the space coordinate. Its presence
introduces all the differences in cosmic evolution. Like FRW
models our field equations are amenable to exact solution only at
extreme values. We find that at early stage our solution reduces
to inhomogeneous analogue of the Einstein-deSitter type of
solution.\\

 2. In line with current observational findings our
model accounts for early deceleration and late acceleration
showing the desirable phenomenon of \emph{flip}in all the examples
we examined. But as expected here the time of flip depends on
space coordinate also. So flip is not synchronous as in FRW cases,
occuring at different shells at different instants. So flip here
is local, not global. Moreover flip occurs early for larger $`r'$.
So for our spherically symmetric model the outer shells will start
acceleration earlier and this is also a good news \emph{vis a vis}
when posited against the problem of shell crossing singularity
\emph{generically} associated with inhomogeneous models.\\

3. Another interesting situation discussed is the possibility of
bounce of our model from a minimum when the arbitrary function of
integration, $C(R)$ assumes a negative value. The bounce also
shares the inhomogeneous characteristic of our model, the
different shells characterised by r-constant hypersurfaces bounce
at different instants. Moreover we have here taken the original
Chaplygin gas in our analysis but now generalised type of equation
of state~\cite{zhang} are being increasingly used with greater
freedom. In our future work we try to extend this work with these
modified Chaplygin gas equations.\\

4. To end the section a final remark may be reemphasised regarding
the apparent accelerated expansion of the universe. To explain the
SNIa observations the concept of accelerated expansion of the
universe need to be invoked only for a FRW type of model. But one
should point out that the Luminosity distance- Redshift relation,
not the accelerated expansion is the quantity that can be directly
measured. And within inhomogeneous models one gets better fit
without the need to introduce the local accelerated expansion and
consequent hypothesis of any extraneous, unphysical matter field
with large negative pressure.\\

\textbf{Acknowledgments}

 \vspace{0.1 cm}
 DP acknowledges financial support
of ERO, UGC  for a Minor Research Project. The financial support
of UGC, New Delhi in the form of a MRP award as also a Twas
Associateship award, Trieste is  acknowledged by SC.

\end{document}